\documentclass[runningheads]{llncs}
\usepackage[T1]{fontenc}
\usepackage{graphicx}
\begin{document}
\title{\textit{AI Expert Twin}: Capturing Expert Cognition for Human-Centred, Practice-Based Learning}
\titlerunning{AI Expert Twin}
\author{Annie Yuan\inst{1}\orcidID{0009-0004-1760-0149} \and
Xiaohua Chen\inst{2}\orcidID{0009-0007-6413-367X} \and
\\ Kalina Yacef \inst{1}\orcidID{0000-0001-7521-6429} \and Judy Kay \inst{1}\orcidID{0000-0001-6728-2768}}
\authorrunning{A. Yuan et al.}
\institute{The University of Sydney, NSW 2006, Australia \and
Beijing Institute of Fashion Technology, No. 2 Yinghua Road, Chaoyang District, Beijing, China, 100029
\\
\email{\{annie.yuan,kalina.yacef,judy.kay\}@sydney.edu.au, chenxh@bift.edu.cn}}

\maketitle              % typeset the header of the contribution
\begin{abstract}
Tacit knowledge embedded in expert practice remains difficult to capture, formalise, and scale. While AI-driven educational systems have advanced personalisation, learner modelling, affective support, and self-regulated learning, they less often model the tacit reasoning and context-sensitive judgement that underpin expert practice in practice-based domains. This paper introduces the \textit{AI Expert Twin}, a cognition-centric framework that models expert knowledge as structured, computable representations of procedural actions, semantic concepts, and decision processes. The framework also considers how value-laden preferences, trade-offs, and uncertainty shape expert judgement in practice. We formalise expert cognition as a three-layer representation and capture knowledge from experts under this model, laying the groundwork for integration into AI-powered educational system. A case study in a cultural heritage workshop demonstrates the feasibility of the approach in a real-world setting. The framework is designed to be transferable across domains such as vocational education and creative industries. By embedding expert heuristics into AI while maintaining transparency and learner agency, the \textit{AI Expert Twin} offers a novel path towards scalable, practice‑based learning and invites further research on ethical, human‑centred applications of AI in education.

\end{abstract}
\section{Introduction}
Artificial intelligence (AI) promises to personalise and democratise learning, but it struggles to represent the nuanced know‑how that experts wield in real contexts. “We can know more than we can tell” captures the enduring challenge of tacit knowledge, the experiential and often intuitive understanding that guides expert action but resists codification \cite{khalili2025unveiling}. Tacit knowing is embodied and context‑specific, extracting it through apprenticeship or mentoring is labour‑intensive and does not scale. Even advanced AI techniques like natural language processing (NLP) and machine learning (ML) can identify patterns, but they cannot preserve the “living context” of tacit expertise.

This gap is especially salient in practice-based domains, where expertise is developed through repeated participation, observation, and situated judgement rather than formal instruction alone. Experienced engineers, clinicians, and artisans often draw on heuristics accumulated over years of practice, recognising patterns, exceptions, and material cues that are difficult to capture in manuals. While AIED has produced sophisticated models of domain knowledge, learner state, and aspects of cognitive skill acquisition, less attention has been given to modelling situated expert judgement as a pedagogical resource. Our focus is therefore not expert performance as an automated decision system, but practice-based expertise that is embodied, tacit, and difficult to articulate as formal rules or solution steps. 

This paper proposes the \textit{AI Expert Twin}: a framework for modelling expert cognition as structured, computable representations of procedural actions, semantic concepts, and decision processes. Our focus is not on modelling expert performance as an automated decision system, but on representing expert reasoning as a pedagogical resource for learners. We outline an initial instantiation through the capture of expert knowledge in a cultural-heritage workshop, demonstrating feasibility and laying the groundwork for integration into AI-powered educational systems. By embedding expert heuristics into AI, the framework offers a new direction for modelling and supporting expert-informed learning.

\section{Related Work}
%\subsection{Knowledge Modelling and Learner Modelling}
Tacit knowledge has long been recognised as central to learning from expert practice. In educational settings, such knowledge is often transmitted through apprenticeship, demonstration, observation, and guided participation. Recent work examine how generative AI might make tacit knowledge more explicit in apprenticeship systems \cite{guo2025making}, and how information systems can document tacit knowledge in traditional handicrafts \cite{yang2024exploring}. However, the challenge remains to represent such knowledge in ways that are pedagogically usable rather than merely documented.

AIED research has produced sophisticated knowledge- and learner-modelling techniques that personalise instruction and have helped scale intelligent tutoring systems (ITS) \cite{vanlehn2013student,bull2010open}. Typical ITSs comprise a domain or an expert model, a student model, a pedagogical model, and an interface \cite{corbett1997intelligent}. In this context, the expert model usually refers to the domain or knowledge model: the skills, concepts, strategies, correct solutions, and misconceptions used to teach and assess students \cite{aleven2023domain,jackson1985introduction}. Cognitive tutors and related systems have also modelled aspects of human problem-solving and skill acquisition \cite{anderson1995cognitive}. Our focus is complementary: modelling the situated heuristics, contextual cues, and judgement processes experts use in practice-based domains.

Digital twin technology provides a useful conceptual foundation: a digital model can mirror a physical system and update over time \cite{tao2022digital}. Cognitive digital twins extend this idea to people by modelling knowledge change and maintaining a ``cognitive thread'' of skill development \cite{saracco2019applying}. However, digital twins focus on tracking state and performance, they do not explicitly model the deeper cognitive processes behind expert decision‑making.

\section{AI Expert Twin Framework}

The \textit{\textit{AI Expert Twin}} is proposed as a framework for capturing, structuring, and operationalising expert cognition for practice-based learning. Its central aim is to provide a representational pathway from expert practice to learner-facing guidance, rather than to assume that tacit expertise can be automatically extracted and tutored at scale. The framework models the transformation from expert practice to learner guidance as:

\begin{center}
\textit{Expert Practice} $\rightarrow$ \textit{Multimodal Data} $\rightarrow$ \textit{Annotated Episodes} $\rightarrow$ \textit{Cognitive Model / Expert Twin} $\rightarrow$ \textit{Learner-Facing Guidance}.
\end{center}

Expert practice is first captured through multimodal data, such as video recordings, expert demonstrations, interviews, annotations, observational notes, and artefact analysis. These data are then segmented into meaningful task episodes. Each episode can be annotated for visible actions, relevant concepts, contextual cues, decision points, and expert explanations. AI tools may assist this process through transcription, video segmentation, multimodal summarisation, and candidate concept extraction; however, expert review remains necessary to validate whether the annotations reflect authentic practice. These annotated episodes are then organised into a structured cognitive model, or \textit{Expert Twin}, that represents how experts act, interpret situations, and make decisions in context. 
% This model then informs an AI tutor that provides learners with adaptive, context-aware guidance and feedback. The learner's interactions and responses can, in turn, generate feedback signals for refining the tutor, the learner model, and, where appropriate, the expert twin itself. 

\begin{figure}[t]
    \centering
    % Replace the filename below with your actual figure file name.
    \includegraphics[width=\linewidth]{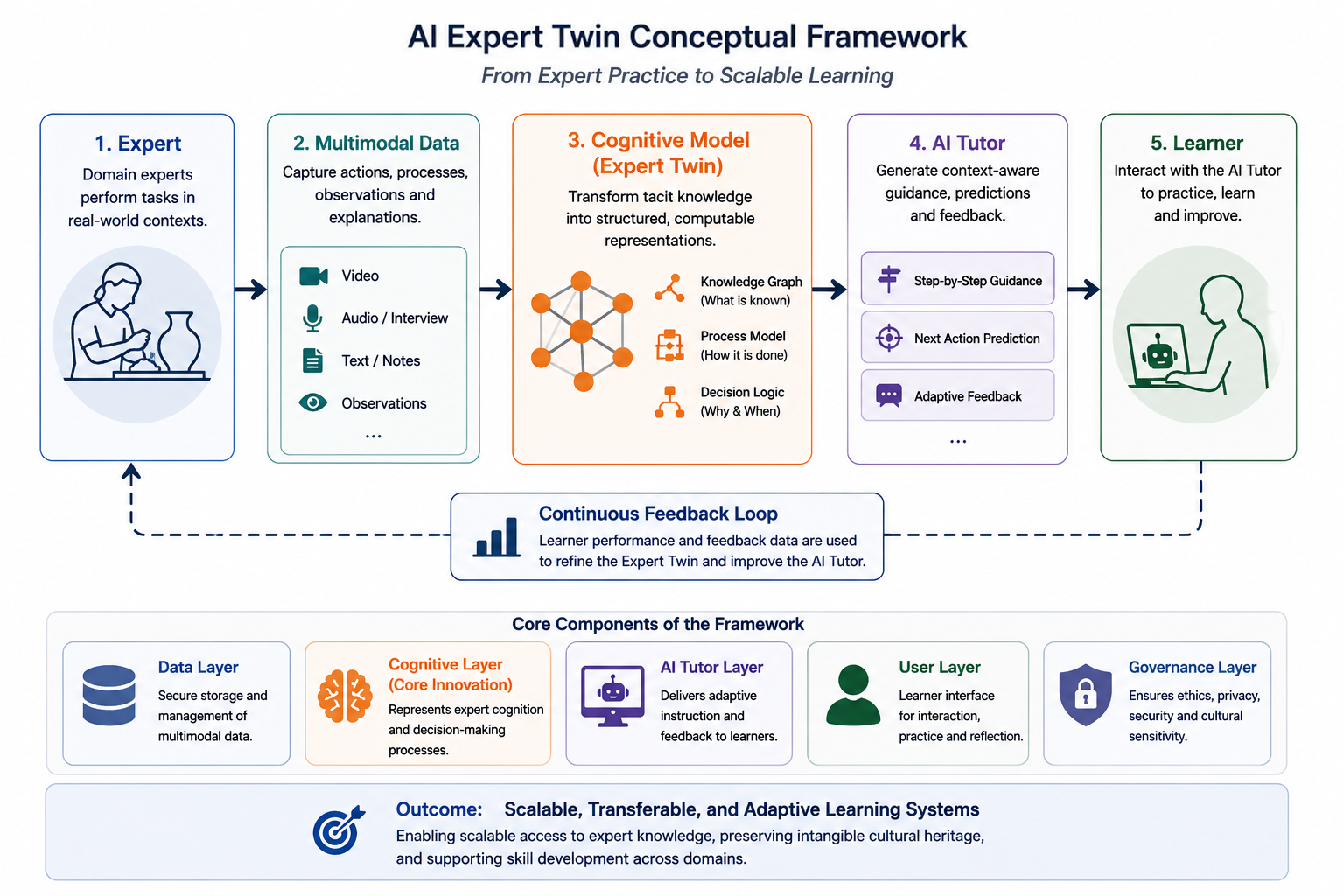}
    \caption{\textit{AI Expert Twin} conceptual framework. Expert practice is captured as multimodal data and transformed into a cognitive model of expert cognition. This Expert Twin informs an AI tutor that supports learners through adaptive guidance and feedback. The framework is organised around five interacting layers: the data layer, cognitive layer, AI tutor layer, user layer, and governance layer.}
\end{figure}

The framework consists of five interacting layers. The \textit{data layer} captures expert practice from real-world activity. This includes multimodal evidence of expert action, such as demonstrations, interviews, observations, annotations, and task artefacts. The purpose of this layer is not simply to record what an expert does, but to preserve the situated context in which expert action occurs.

The \textit{cognitive layer} is the core contribution of the \textit{AI Expert Twin}. It transforms annotated episodes into structured representations of how experts think, decide, and act. This layer models expert cognition through three interrelated components: a procedural layer, a semantic layer, and a decision layer. Within the decision layer, we also consider value-laden preferences, trade-offs, and uncertainty as factors that shape expert judgement in practice. Together, these components form a computable representation of expert practice.

The \textit{AI tutor layer} uses the Expert Twin as a structured knowledge base for learner-facing support. The procedural layer can inform step-by-step scaffolding and workflow guidance; the semantic layer can support explanations of concepts, materials, tools, and relationships; and the decision layer can support reflective prompts, cue-based feedback, and explanations of why an expert might choose one action over another in context. The \textit{user layer} represents learners' engagement with the tutor, including their questions, actions, reflections, errors, and progress. Finally, the \textit{governance layer} cuts across the whole framework. It addresses transparency, learner agency, data privacy, expert consent, cultural sensitivity, and the risk of treating one expert's practice as the only valid model of expertise.

\subsection{Three-Layer Representation of Expert Cognition}

We define the \textit{AI Expert Twin} as a structured, computable representation of expert cognition designed to support AI-mediated practice-based learning. The cognitive layer consists of three interrelated components.

\textbf{Procedural layer.} The procedural layer captures what experts do. E.g., task decomposition, workflows, dependencies, and variations in practice.

\textbf{Semantic layer.} The semantic layer captures what experts know and attend to: concepts, entities, materials, tools, constraints, and relationships, which may be represented through knowledge graphs or other structured models.

\begin{figure}[t]
    \centering
    \includegraphics[width=\linewidth]{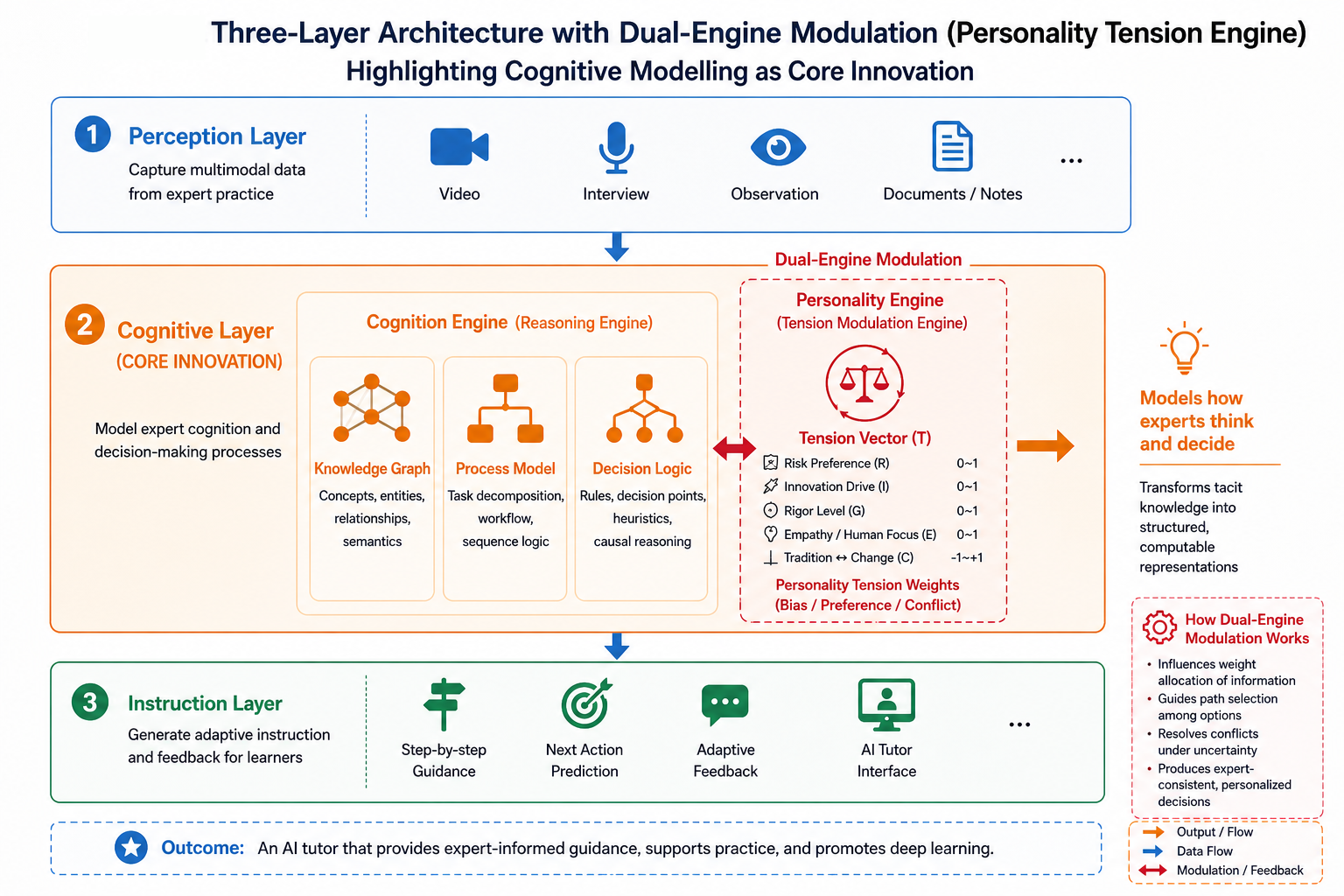}
    \caption{Three-layer representation of expert cognition in the AI Expert Twin framework, with a personality/tension engine modulating the decision layer through tension vector $T$.}
    \vspace{-12pt} 
\label{fig:three-layer-architecture}
\end{figure}

\textbf{Decision layer.} The decision layer captures expert judgement as context-sensitive choice. It encodes rules, heuristics, cues, trade-offs, and decision strategies, then applies a personality/tension vector $T=\{R,I,G,E,C\}$, where $R$ denotes risk preference, $I$ innovation drive, $G$ rigor, $E$ empathy or human focus, and $C$ orientation toward tradition or change. These weights influence how experts select among candidate actions and explain decisions under uncertainty.

Together, these three layers form a computable model of expert cognition. The procedural layer captures action, the semantic layer captures meaning, and the decision layer captures judgement. In this way, the \textit{AI Expert Twin} connects expert practice with AI-supported learning interfaces that can explain, scaffold, and adapt guidance for learners.

%\subsection{From Expert Practice to Learner Guidance}

The \textit{AI Expert Twin} acts as a bridge between expert practice and learner interaction. Rather than replacing experts or automating judgement, it makes selected aspects of expert cognition visible and pedagogically usable. In a learning system, the procedural layer can support workflow scaffolding, the semantic layer can support conceptual explanations, and the decision layer can support reflective prompts or decision rationales. The personality/tension engine further represents the value-laden preferences, trade-offs, and pedagogical stance that shape an expert's approach to practice. Thus, future educational systems could allow learners to engage not only with different learning content, but also with different expert perspectives, such as more cautious, tradition-oriented, experimental, rigorous, or human-centred approaches to the same task. For example, the tutor could explain which contextual cues a particular expert would attend to, why that expert would judge an action as appropriate, and how another expert might make a different yet defensible decision.

\section{Initial Instantiation: Cultural Heritage Workshop}

We ground the AI Expert Twin framework in an ongoing cultural heritage collaboration focused on jade carving. The workshop involved jade carving master Tiecheng Zhang and a research team led by Professor Xiaohua Chen, who conducted interviews and captured multimodal records of the master’s practice. Rather than presenting a completed AI tutor evaluation, this section illustrates the early-stage process of collecting and structuring expert practice as a precursor to AI-powered learning support.

\begin{figure}[t]
    \centering
    \includegraphics[width=\linewidth]{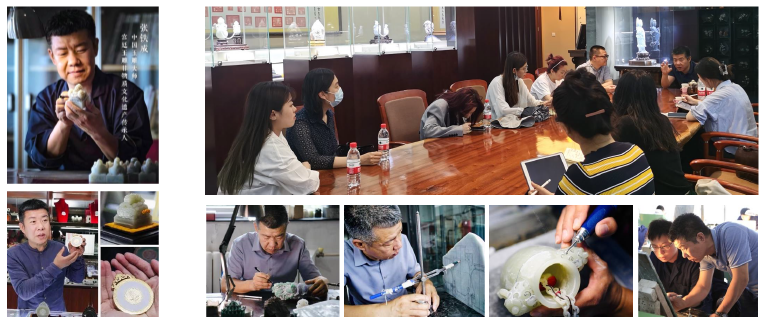}
    \caption{Jade carving workshop and multimodal data capture. Left: jade carving master Tiecheng Zhang and selected signature works, including the Beijing Olympics jade-inlaid gold medal. Right: interviews conducted by the research team led by Professor Xiaohua Chen, and multimodal records of the master's workflow from design and carving to apprenticeship-based instruction.}
    \label{fig:jade-carving}
\end{figure}

As shown in Figure~\ref{fig:jade-carving}, the data capture included interviews with the master, documentation of signature artefacts, and multimedia records of the workflow from design to carving and apprenticeship-based instruction. These records provide evidence of expert action, explanation, material judgement, and interaction with learners or apprentices. In the current stage, the data are being used to identify meaningful task episodes and to examine how expert knowledge can be represented through the three-layer model.

The three-layer model provides a structure for organising this knowledge. The procedural layer captures visible sequences of practice, such as design preparation, tool selection, carving actions, and demonstration steps. The semantic layer captures the concepts, materials, techniques, motifs, artefact features, and quality criteria that give meaning to those actions. The decision layer captures expert heuristics, such as how the master judges material properties, recognises risk, identifies errors, adjusts pressure or technique, and decides when to continue, pause, or revise a process. These decision episodes may also reveal value-laden tensions, such as balancing preservation with innovation, or caution with expressive risk-taking.

This case shows the feasibility of using cultural heritage practice as a site for capturing expert cognition, while marking the limits of the current work, we have not yet completed a full Expert Twin or evaluated an AI tutor with learners. Future work will refine annotation, validate the model with experts, and investigate how it supports feedback, reflection, and guided practice.

\section{Implications for Next-Generation Learning Interfaces}

The AI Expert Twin suggests that next-generation learning interfaces should personalise not only to learner state, but also around inspectable expert reasoning. It relates to prior work on educational agents, learning companions, and AI companions, which examined how artificial peers, tutors, or companions can support learning through interaction \cite{chou2003redefining,chou2025defining}. Our focus is different, rather than designing an agent persona, we propose a representational structure for exposing expert practice as procedural, semantic, decision-making, and value-laden knowledge.  

The interface does not need to simulate a complete autonomous expert. Instead, the Expert Twin serves as a structured reference for prompts, explanations, and feedback. In jade carving, for example, a learner may follow the correct sequence but hesitate when stone texture changes. An Expert Twin-based interface could identify the task stage from the procedural layer, explain material properties from the semantic layer, and use the decision layer to highlight expert cues such as resistance, translucency, symmetry, or fracture risk.

This enables interaction around expert reasoning and context, not only task correctness. In the case study of jade carving, if a learner is unsure whether to pause and inspect the stone, the interface might prompt them to examine the relevant cues: \textit{Look at the surface texture and resistance. What do these suggest about the risk of continuing?} It might then explain: \textit{Master Zhang recommends slowing down here because a change in resistance can indicate higher fracture risk.'} Where multiple actions are defensible, the personality tension vector can show how value-laden preferences shape judgement, such as caution versus expressive risk-taking or preservation versus innovation. Learners can therefore inspect not only what an expert might do but also why different experts may make different defensible decisions.  

This connects expert modelling with learner modelling: learner models represent what a learner knows, does, or struggles with, while the Expert Twin represents target patterns of expert action, meaning-making, judgement, and tension-weighted decision-making. Comparing these models could support feedback on whether learners attend to appropriate cues, apply relevant concepts, and make decisions at suitable moments, without enforcing a single correct path. In this sense, the AI tutor is not merely a source of answers, but an interface through which learners interact with structured expert cognition, mediating between learners and expert practice rather than replacing teachers, experts, or learning companions.

\section{Conclusion}

This position paper argues that AI-powered learning systems should move beyond modelling domain content and learner state to also model expert cognition as a pedagogically usable resource. The \textit{AI Expert Twin} offers a framework for structuring expert practice through procedural, semantic, and decision layers, making expert reasoning more visible and interactive for practice-based learning.

\begin{credits}
\subsubsection{\ackname} We thank Master Tiecheng Zhang for sharing his time, expertise and practice. This research is supported by the University of Sydney Postgraduate Award. 

\subsubsection{\discintname}
The authors have no competing interests to declare that are
relevant to the content of this article. 
\end{credits}
\bibliographystyle{splncs04}
\bibliography{bibliography}

\end{document}